\documentclass{ws-procs9x6}

\def\bea{\begin{eqnarray}}
\def\eea{\end{eqnarray}}

\newcommand\lsim {~^{<~}_{\sim~}}

\begin{document}

\title{Quark Confinement Physics from Lattice QCD
\vspace{-0.5cm}}

\author{H. Suganuma, K. Amemiya, H. Ichie, H. Matsufuru, Y. Nemoto and T.T. Takahashi}

\address{Research Center for Nuclear Physics, Osaka University,\\
Mihogaoka 10-1, Ibaraki  567-0047, Japan
\vspace{-0.3cm}
}
\maketitle
\abstracts{ 
We study quark confinement physics using lattice QCD. In the maximally 
abelian (MA) gauge, the off-diagonal gluon amplitude is strongly suppressed, 
and then the off-diagonal gluon phase shows strong randomness, 
which leads to  a large effective off-diagonal gluon mass, 
$M_{\rm off} \simeq 1.2 {\rm GeV}$.
Due to the large off-diagonal gluon mass in the MA gauge,  
low-energy QCD is abelianized like nonabelian Higgs theories.
In the MA gauge, there appears a macroscopic network of the monopole 
world-line covering the whole system. 
We extract and analyze the dual gluon field $B_\mu$ from 
the monopole-current system in the MA gauge, and  
evaluate the dual gluon mass as $m_B = 0.4 \sim$ 0.5GeV 
in the infrared region, which is a lattice-QCD evidence of 
the dual Higgs mechanism by monopole condensation. 
Even without explicit use of gauge fixing, we can define the maximal 
abelian projection by introducing a ``gluonic Higgs field'' $\phi(x)$, 
whose hedgehog singularities lead to monopoles. 
From infrared abelian dominance and infrared monopole condensation, 
infrared QCD is describable with the dual Ginzburg-Landau theory. 
In relation to the color-flux-tube picture for baryons, 
we study the three-quark (3Q) ground-state potential $V_{\rm 3Q}$ 
in SU(3)$_c$ lattice QCD at the quenched level, 
with the smearing technique for enhancement of the ground-state component.
With accuracy better than a few \%, 
$V_{\rm 3Q}$ is well described by a sum of 
the two-body Coulomb part 
and the three-body linear confinement part $\sigma_{\rm 3Q} L_{\rm min}$, 
where $L_{\rm min}$ denotes the minimal value of the total 
length of the color flux tube linking the three quarks. 
Comparing with the Q-$\bar {\rm Q}$ potential, 
we find a universal feature of the string tension as 
$\sigma_{\rm 3Q} \simeq \sigma_{\rm Q \bar Q}$ 
and the OGE result for the Coulomb coefficient as 
$A_{\rm 3Q} \simeq \frac12 A_{\rm Q \bar Q}$. 
\vspace{-0.3cm}
} 

\section{Nonabelian Feature in QCD}

Nowadays, quantum chromodynamics (QCD) is established 
as the fundamental theory of the strong interaction. 
In spite of the simple form of QCD lagrangian 
\bea
{\mathcal L}_{\rm QCD}=-{1 \over 2} {\rm tr} G_{\mu\nu}G^{\mu\nu}
+\bar q (i \hspace{0.0cm}{\not \hspace{-0.1cm} D}-m_q) q, 
\eea
it is still hard to understand nonperturbative QCD (NP-QCD) phenomena,  
such as color confinement and dynamical chiral-symmetry breaking,   
in the infrared strong-coupling region. 
In fact, QCD tells us the ``rule'' of the elementary interaction 
between quarks and gluons, but to solve QCD is another difficult problem.

In this point, QCD may resemble the Rubik Cube, where the
rule is quite simple, but to solve this puzzle is rather difficult.
Like QCD, the difficulty of the Rubik cube comes from the non-commutable
procedures of the ``nonabelian'' rotational process as shown in Fig.1.
Here, a kind of ``local procedure'' on 3 $\times$ 3 $\times$ 3 
subcubes leads to extremely large number of configurations,
which makes this puzzle difficult and interesting. 
Then, the Rubik cube may be regarded as a miniature of QCD.\cite{NEWS99}
\begin{figure}[hb]
\begin{center}
\includegraphics[height=7cm]{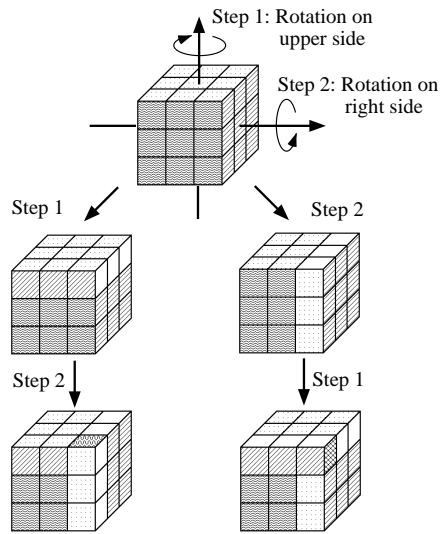}
\caption{
The difficulty and the interest of the Rubik cube originate from
non-commutable operations based on the nonabelian feature 
of the rotational group. The configuration after Step 1 and Step 2
depends on the order of these operations.}
\end{center}
\end{figure}
The main difficulties of QCD originate from the nonabelian  
and the strong-coupling features in the infrared region below 1 GeV.
In the ultraviolet region, the QCD coupling is weak and then 
we can use the perturbative QCD, where  
the three and the four gluon vertices stemming from 
the nonabelian nature of QCD are treated 
as perturbative interactions. 
In the infrared region, however, the nonabelian and
the strong-coupling features are significant. 

In the electro-magnetism, the superposition of solutions is possible, 
because of the linearity of the field equation 
$
\partial_\mu F^{\mu\nu}=j^\nu, 
$
so that we can consider the partial electro-magnetic field
formed by each charge and the total electro-magnetic field can be
obtained by summing up the individual field solution. 
On the other hand, 
the QCD field equation becomes nonlinear as 
\bea
\partial_\mu G^{\mu\nu} + ig [A_\mu, G^{\mu\nu}]=j^\nu,
\eea
due to the nonabelian feature, 
and then it is difficult to solve it even at the classical level,
because the superposition of solutions is impossible. 
So, we cannot divide the color-electromagnetic field into
each part formed by each quark. Instead, the QCD system is
to be analyzed as a whole system.
In this way, for the analysis of QCD, the nonabelian feature provides 
one of serious difficulties. 

\vspace{-0.25cm}

\section{Abelianization of QCD in MA Gauge}

The nonabelian nature of QCD in the infrared region is, however, removed 
in the MA gauge. 
In fact, QCD is reduced into an abelian gauge theory 
with color-magnetic monopoles, keeping essence of 
infrared nonperturbative features.

\vspace{-0.25cm}

\subsection{Dual Superconductor Theory and QCD}

In QCD, to understand the confinement mechanism is 
one of the most difficult problems remaining in the particle physics.
As is indicated by hadron Regge trajectories and lattice QCD 
calculations, the confinement phenomenon is 
characterized by {\it one-dimensional squeezing} of the 
color-electric flux and 
the {\it string tension} $\sigma \simeq 1{\rm GeV/fm}$, 
which is the key quantity of confinement.

On the confinement mechanism, 
Nambu first proposed the {\it dual superconductor theory} for quark 
confinement,\cite{N74} based on the electro-magnetic duality in 1974.  
In this theory, there occurs the one-dimensional squeezing 
of the color-electric flux between quark and anti-quark 
by the {\it dual Meissner effect} due to condensation of 
bosonic color-magnetic monopoles. 
However, there are {\it two large gaps} between QCD and the 
dual superconductor theory.$^{3-5}$
\begin{enumerate} 
\item 
The dual superconductor theory is based on the {\it abelian gauge theory} 
subject to the Maxwell-type equations, where electro-magnetic duality is 
manifest, while QCD is a nonabelian gauge theory.   
\item
The dual superconductor theory requires condensation of 
color-magnetic monopoles as the key concept, while QCD does not 
have color-magnetic monopoles as the elementary degrees of freedom.
\end{enumerate}
These gaps may be simultaneously fulfilled by taking 
{\it MA gauge fixing,} which reduces QCD to an abelian gauge theory 
including color-magnetic monopoles.

\subsection{MA Gauge Fixing and Relevant Mode for Confinement}

In Euclidean QCD, the maximally abelian (MA) gauge is 
defined so as to minimize the total amount of the 
off-diagonal gluons,$^{3-5}$ 
\begin{eqnarray}
R_{\rm off} [A_\mu ( \cdot )] \equiv \int d^4x \ {\rm tr}
\left\{ 
[\hat D_\mu ,\vec H][\hat D_\mu ,\vec H]^\dagger 
\right\} 
={e^2 \over 2} \int d^4x \sum_\alpha |A_\mu ^\alpha (x)|^2, 
\end{eqnarray}
by the SU($N_c$) gauge transformation.
Here, we have used the Cartan decomposition, 
$A_\mu (x)=\vec A_\mu (x) \cdot \vec H 
+\sum_\alpha A_\mu^\alpha (x)E^\alpha $. 
Since the ${\rm SU}(N_c)$ covariant derivative operator 
$\hat D_\mu \equiv \hat \partial_\mu+ieA_\mu $ obeys the 
adjoint gauge transformation, the local form of 
the MA gauge condition is easily derived as 
$[\vec H, [\hat D_\mu , [\hat D_\mu , \vec H]]]=0.$ $^{3-5}$ 

In the MA gauge, the gauge symmetry 
$G \equiv {\rm SU}(N_c)_{\rm local}$ 
is reduced into $H \equiv {\rm U(1)}_{\rm local}^{N_c-1} 
\times {\rm Weyl}^{\rm global}_{N_c}$, 
where the global Weyl symmetry is a subgroup of ${\rm SU}(N_c)$ 
relating the permutation of $N_c$ bases in the fundamental 
representation. 
In the MA gauge, off-diagonal gluons behave as charged matter fields like 
$W_\mu^{\pm}$ in the Standard Model, and provide the 
color-electric current in terms of the residual abelian gauge symmetry. 
In addition, according to the reduction of the gauge symmetry, 
color-magnetic monopoles appear as topological objects 
reflecting the nontrivial homotopy group$^{3-8}$
\bea
\Pi_2({\rm SU}(N_c)/{\rm U(1)}^{N_c-1})=\Pi_1({\rm U(1)}^{N_c-1})
={\bf Z}^{N_c-1}_\infty, 
\eea
in a similar manner to similarly in the GUT monopole.
Here, the global Weyl symmetry and color-magnetic monopoles are 
relics of nonabelian nature of QCD. 

Thus, in the MA gauge, QCD is reduced into an abelian gauge theory 
including color-magnetic monopoles, 
which is expected to provide a theoretical basis of the 
dual superconductor theory for quark confinement.
Furthermore, recent lattice QCD studies show remarkable features of 
{\it abelian dominance} and {\it monopole dominance} for NP-QCD 
in the MA gauge.

\begin{enumerate}
\item
Without gauge fixing, all the gluon components equally contribute to 
NP-QCD, and it is difficult to extract relevant degrees of freedom for NP-QCD. 
\item
In the MA gauge, QCD is reduced into an abelian gauge theory 
including the electric current $j_\mu $ and the magnetic current $k_\mu $, 
which forms the 
{\it global network of the monopole world-line covering the whole system.} 
In the MA gauge, lattice QCD shows {\it abelian dominance} 
for NP-QCD (confinement, chiral symmetry breaking \cite{M95},
gluon propagators \cite{AS99}): only the diagonal gluon, 
which remains an abelian gauge field, 
is relevant for NP-QCD, while off-diagonal gluons do not 
contribute to NP-QCD. 
\item
By the Hodge decomposition, the diagonal gluon is decomposed into  
the ``photon part'' and the ``monopole part'', 
corresponding to the separation of $j_\mu$ and $k_\mu$. 
In the MA gauge, lattice QCD shows 
{\it monopole dominance}$^{9,11-12}$
for NP-QCD: the monopole part ($k_\mu \ne 0$, $j_\mu=0$) leads to NP-QCD, 
while the photon part ($j_\mu \ne 0, k_\mu=0$) seems trivial like QED 
and does not contribute to NP-QCD. 
For example, on the $Q$-$\bar Q$ potential, 
the purely linear confinement potential appears in the monopole part, 
while the Coulomb potential appears in the photon part 
like QED.\cite{P97}
\end{enumerate}
In fact, by taking the MA gauge, the {\it relevant collective mode for NP-QCD} 
can be extracted as the color-magnetic monopole.\cite{SAIT00,SITA98} 

\vspace{-0.3cm}

\section{Strong Random Phase of Off-diagonal Gluon in MAQCD}

To find out essence of the MA gauge,  
we study the feature of the off-diagonal gluon field 
$A_\mu^\pm \equiv \frac1{\sqrt{2}}(A_\mu^1 \pm i A_\mu^2)$ 
in the MA gauge in SU(2) lattice QCD. 

In SU(2) lattice QCD, the SU(2) link variable is factorized
as $U_\mu(s)=M_\mu(s)u_\mu(s)$, according to the Cartan decomposition
${\rm SU(2)/U(1)}_3 \times {\rm U(1)}_3$.
Here, $u_\mu(s)\equiv \exp\{i\tau^3\theta^3_\mu(s)\}\in {\rm U(1)}_3$
denotes the abelian link variable, and the abelian projection is defined 
by the replacement as $U_\mu(s) \rightarrow u_\mu(s)$. 
The off-diagonal matrix 
$M_\mu(s)\in {\rm SU(2)/U(1)}_3$ is parameterized as
\begin{eqnarray}
M_\mu(s)\equiv e^{i\{\tau^1\theta^1_\mu(s)+\tau^2\theta^2_\mu(s)\}}
=\left(
\begin{array}{cc}
{\rm cos}{\theta_\mu}(s) & i e^{-i\chi_\mu(s)}{\rm sin}{\theta_\mu}(s)  \\
i  e^{i\chi_\mu(s)}{\rm sin}{\theta_\mu}(s) & {\rm cos}{\theta_\mu}(s)
\end{array}
\right).
\end{eqnarray}
In the continuum limit,  
$\chi_\mu(s)$ coincides with the off-diagonal gluon phase as 
$A_\mu^\pm(x)=e^{\pm i \chi_\mu(x)}|A_\mu^\pm(x)|$, 
and $\theta_\mu(s)$ is proportional to the off-diagonal gluon amplitude as 
$\theta_\mu(s)=\frac1{\sqrt{2}}a|eA_\mu^{\pm}(s)|$ 
with the lattice spacing $a$.
In the MA gauge, the diagonal element \ $\cos \theta_\mu(s)$
in $M_\mu(s)$ is maximized by the SU(2) gauge transformation, 
{\it e.g.} $\langle\cos \theta_\mu(s)\rangle_{\rm MA}\simeq 0.93$
at $\beta=2.4$.
Accordingly, the off-diagonal gluon 
$A_\mu^\pm(x)=e^{\pm i \chi_\mu(x)}|A_\mu^\pm(x)|$ 
has two relevant features in the MA gauge.$^{3-5}$
\begin{enumerate}
\item
The off-diagonal gluon amplitude $|A_\mu^{\pm}(x)|$ 
(or $|\sin \theta_\mu(s)|$ on lattices) 
is strongly suppressed by SU($N_c$) gauge transformation in the MA gauge.
\item
The off-diagonal gluon phase $\chi_\mu(x)$ 
tends to be random, because $\chi_\mu(x)$ is not 
constrained by MA gauge fixing at all, 
and only the constraint from the QCD action is weak 
due to a small accompanying factor $|A_\mu^\pm|$ or $|\sin \theta_\mu|$.
\end{enumerate}

\begin{figure}[hb]
\vspace{-0.2cm}
\begin{center}
\includegraphics[height=4cm]{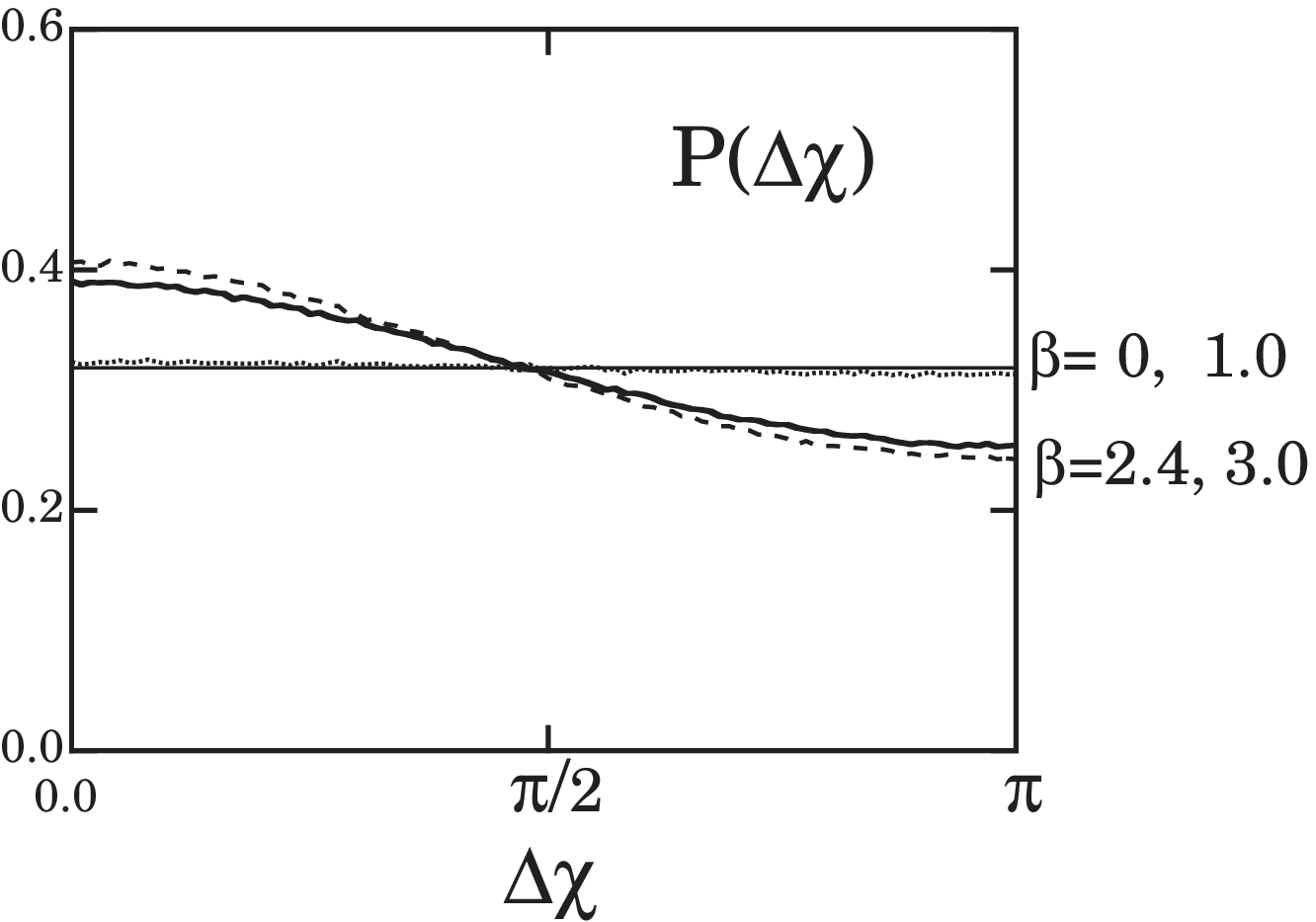}
\includegraphics[height=3.9cm]{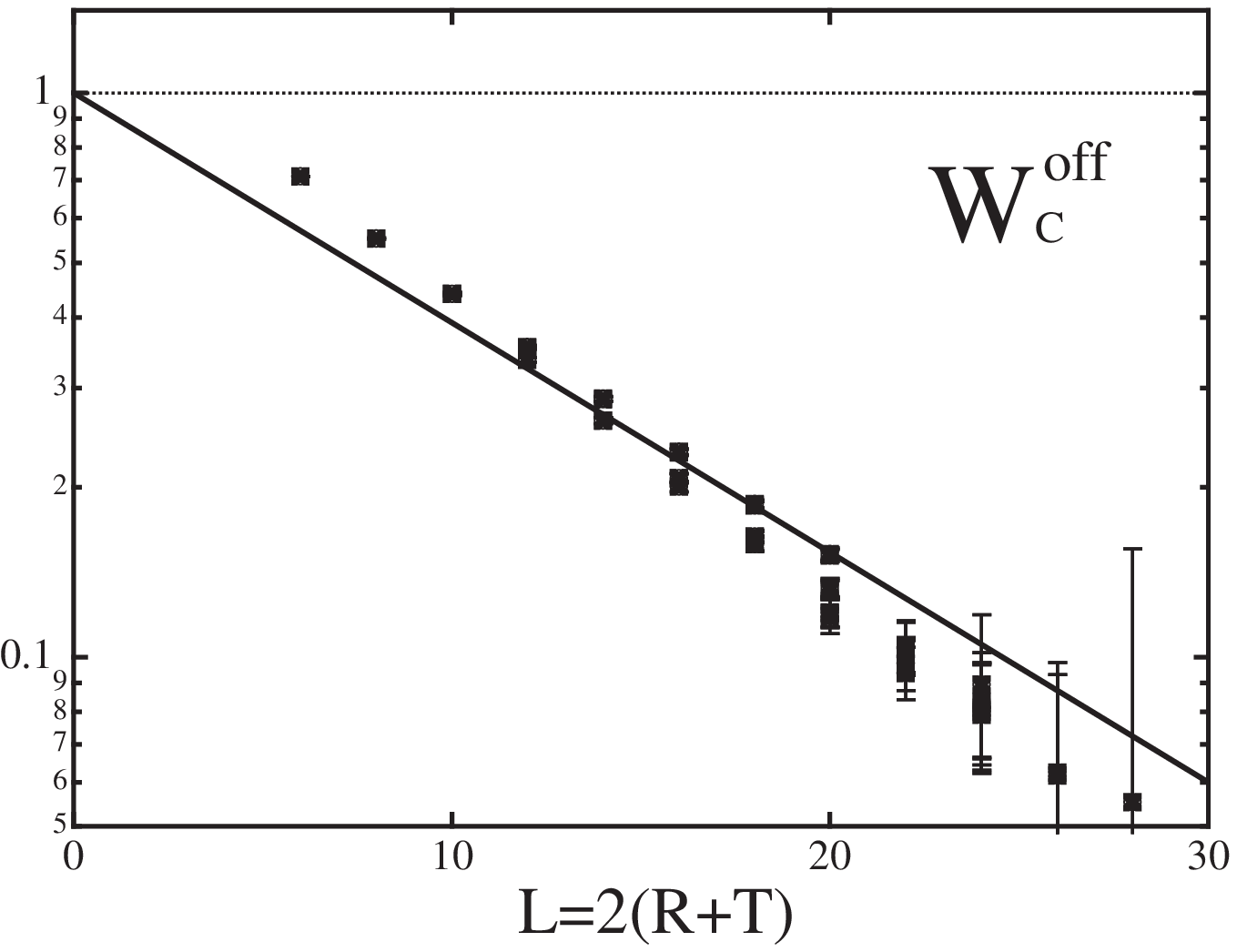}
\caption{
(a) The distribution $P(\Delta \chi)$ of the difference 
$\Delta \chi \equiv |\chi_\mu(s)-\chi_\mu(s+\hat \nu)| ({\rm mod} \pi)$ 
in the MA gauge plus U(1)$_3$ Landau gauge at $\beta$=0 
($a=\infty$, thin line), 
$\beta$=1.0 
($a \simeq$ 0.57fm, dotted curve), 
$\beta$=2.4 ($a \simeq$ 0.127fm, solid curve), 
$\beta$=3.0 ($a \simeq$ 0.04fm, dashed curve).
(b) The off-diagonal gluon contribution to the Wilson loop 
$W_C^{\rm off}$ 
v.s. the perimeter length $L \equiv 2(R+T)$ 
in the MA gauge on $16^4$ lattice with $\beta$=2.4.
The thick line denotes the theoretical estimation in Eq.(10) 
with the microscopic input 
$\langle \ln \{ \cos \theta_\mu(s) \} \rangle_{\rm MA} \simeq -0.082$ 
at $\beta=2.4$.}
\end{center}
\vspace{-0.5cm}
\end{figure}

Now, we consider the behavior of 
$\Delta \chi \equiv |\chi_\mu(s)-\chi_\mu(s+\hat \nu)| ({\rm mod} \pi)$ 
in the MA gauge. 
If the off-diagonal gluon phase  $\chi_\mu(x)$ is a continuum variable,  
as lattice spacing $a$ goes to 0, 
$\Delta \chi \simeq a |\partial_\nu \chi_\mu|$ must 
go to zero, and hence $P(\Delta \chi)$ approaches to 
the $\delta$-function with the peak at $\Delta \chi=0$. 
However, as shown in Fig.2(a), $P(\Delta \chi)$ is 
almost $a$-independent and almost flat. These features
indicate the {\it strong randomness of the off-diagonal gluon phase} 
$\chi_\mu(x)$ in the MA gauge. 
Then, $\chi_\mu(x)$ is approximately a random angle variable 
in the MA gauge.\footnote{
However, as shown in Sect.~6, near the monopole, there remains 
a large amplitude of $|\sin \theta_\mu(s)|$ even in the MA gauge, 
and $\chi_\mu(s)$ is strongly constrained so as to make the QCD action 
small. Hence, $\chi_\mu(s)$ cannot be regarded as a random 
variable near monopoles.}

\vspace{-0.3cm}

\subsection{Randomness of Off-diagonal Gluon Phase and Abelian Dominance} 

Within the random-variable approximation$^{3-5}$ 
for the off-diagonal gluon phase $\chi_\mu(s)$ in the MA gauge, 
we analytically prove abelian dominance of the string tension. 
Here, we use
\begin{eqnarray}
\langle e^{i\chi_\mu(s)}\rangle_{\rm MA} \simeq 
\frac1{2\pi}  \int_0^{2\pi} d\chi_\mu(s)\exp\{i\chi_\mu(s)\}=0. 
\end{eqnarray}
In the Wilson loop $\langle W_C[U]\rangle \equiv
\langle{\rm tr}\Pi_C U_\mu(s) \rangle=
\langle{\rm tr}\Pi_C\{M_\mu(s)u_\mu(s)\}\rangle$,
the off-diagonal matrix $M_\mu(s)$ is reduced to a $c$-number factor, 
\begin{eqnarray}
M_\mu(s) \rightarrow \cos \theta_\mu(s) \ {\bf 1},
\end{eqnarray}
and then the SU(2) link variable $U_\mu(s)$ 
is reduced to a {\it diagonal matrix} as  
\begin{eqnarray}
U_\mu(s)\equiv M_\mu(s)u_\mu(s)
\rightarrow 
\cos \theta_\mu(s) u_\mu(s), 
\end{eqnarray}
after the integration over $\chi_\mu(s)$. 
For the $R \times T$ rectangular $C$, the Wilson loop 
$W_C[U]$ in the MA gauge is approximated as 
\begin{eqnarray}
\langle W_C[U]\rangle
	&=&
	\langle{\rm tr}\Pi_{i=1}^L 
	\{M_{\mu_i}(s_i)u_{\mu_i}(s_i)\}\rangle
	\simeq 
	\langle\Pi_{i=1}^L \cos \theta_{\mu_i}(s_i) \cdot 
	{\rm tr} \Pi_{j=1}^L u_{\mu_j}(s_j)\rangle_{\rm MA} \nonumber \\ 
	 &\simeq& 
	\langle\exp\{\Sigma_{i=1}^L 
		\ln (\cos \theta_{\mu_i}(s_i))\}\rangle_{\rm MA} 
	\ \langle W_C[u]\rangle_{\rm MA} 
\end{eqnarray}
with perimeter $L \equiv 2(R+T)$ and 
the abelian Wilson loop 
$W_C[u] \equiv {\rm tr}\Pi_{i=1}^L u_{\mu_i}(s_i)$. 
Replacing 
$\sum_{i=1}^L \ln \{\cos(\theta_{\mu_i}(s_i))\}$ by its average 
$L \langle \ln (\cos \theta_\mu) \rangle_{\rm MA}$
in a statistical sense, 
we derive the {\it perimeter law} of 
the {\it off-diagonal gluon contribution to the Wilson loop} as 
\bea
W_C^{\rm off}\equiv 
\langle W_C[U]\rangle/\langle W_C[u]\rangle_{\rm MA}
\simeq \exp\{L \langle \ln (\cos \theta_{\mu}) \rangle_{\rm MA}\}, 
\eea
which is confirmed in lattice QCD, as shown in Fig.2(b).
Near the continuum limit, we find also 
the relation between the {\it macroscopic} quantity $W_C^{\rm off}$ 
and the {\it microscopic} quantity of the off-diagonal gluon amplitude  
$\langle |eA_\mu^\pm|^2 \rangle_{\rm MA}$ as 
\bea
W_C^{\rm off}\equiv 
\langle W_C[U]\rangle/\langle W_C[u]\rangle_{\rm MA}
\simeq \exp\{-L \frac{a^2}{4}
\langle |eA_\mu^\pm|^2 \rangle_{\rm MA}\}. 
\eea
In this way, {\it perfect abelian dominance for the string tension},  
$\sigma_{\rm SU(2)}=\sigma_{\rm Abel}$, 
is analytically derived within the random-variable approximation 
for the off-diagonal gluon phase in the MA gauge. 

\subsection{Strongly Random Phase and Large Mass of Off-diagonal Gluons}

As another remarkable fact, 
{\it strong randomness of off-diagonal gluon phases leads to 
rapid reduction of off-diagonal gluon correlations.} 
In fact, if $\chi_\mu(x)$ is a complete random phase, 
Euclidean off-diagonal gluon propagators exhibit 
$\delta$-functional reduction,
\bea
\langle A_\mu^+(x) A_\nu^-(y) \rangle_{\rm MA} 
&=&
\langle |A_\mu^+(x)||A_\nu^-(y)|e^{i\{\chi_\mu(x)-\chi_\nu(y)\}} 
\rangle_{\rm MA} \nonumber \\
&=&
\langle |A_\mu^\pm(x)|^2 \rangle_{\rm MA} \delta_{\mu\nu}\delta^4(x-y), 
\eea
which means the infinitely large mass of off-diagonal gluons.
Of course, the real off-diagonal gluon phases are not complete 
but approximate random phases. 
Then, off-diagonal gluon mass would be large but finite. 
In this way, {\it strong randomness of off-diagonal gluon phases 
is expected to provide a large effective mass of off-diagonal gluons.} 

\section{Large Mass Generation of Off-diagonal Gluons in MA Gauge : Essence of Infrared Abelian Dominance}

Using SU(2) lattice QCD, 
we actually investigate the Euclidean gluon propagator 
$G_{\mu \nu }^{ab} (x-y) \equiv \langle A_\mu ^a(x)A_\nu ^b(y)\rangle$ 
($a,b =1,2,3$) and the off-diagonal gluon mass $M_{\rm off}$ 
in the MA gauge. As for the residual U(1)$_3$ gauge symmetry, 
we take U(1)$_3$ Landau gauge, 
to extract most continuous gluon configuration under 
the MA gauge constraint and to compare with the continuum theory.
The continuum gluon field $A_\mu^a(x)$ is derived from 
the link variable as 
$U_\mu(s)={\rm exp}\{iaeA_\mu^a(s) \frac{\tau^a}{2}\}$.

We show in Fig.3(a) the scalar-type gluon propagators 
$G_{\mu \mu}^3(r)$ and 
$G_{\mu\mu}^{+-}(r) \equiv \langle A_\mu^{+}(x)A_\mu^{-}(y)\rangle
= \frac12 \{G_{\mu\mu}^1(r)+G_{\mu\mu}^2(r)\}$, 
which depend only on the four-dimensional Euclidean 
distance $r \equiv \sqrt{(x_\mu- y_\mu)^2}$, 
in SU(2) lattice QCD 
with $2.2 \le \beta \le 2.4$ with various sizes 
($12^3 \times 24$, $16^4$, $20^4$).
We find {\it infrared abelian dominance for the gluon propagator 
in the MA gauge}: only the abelian gluon $A_\mu^3(x)$ propagates over 
the long distance and can influence the long-distance 
physics.$^{4,5,10}$

Since the four-dimensional Euclidean propagator of the 
massive vector boson with the mass $M$ takes a 
Yukawa-type asymptotic form as 
\bea
G_{\mu\mu}(r) = \frac3{4\pi^2} \frac{M}{r} K_1(Mr)+\frac1{M^2}\delta^4(x-y)
\simeq \frac{3M^{1/2}}{2(2\pi)^{3/2}}\frac{e^{-Mr}}{r^{3/2}},
\eea
we investigate the effective mass $M_{\rm off}$ of off-diagonal gluons 
$A_\mu^{\pm}(x)$ from the slope of the logarithmic plot of 
$r^{3/2} G_{\mu\mu}^{+-}(r)\sim \exp\{-M_{\rm off}r\}$ in Fig.3(b). 
\begin{enumerate}
\item
The off-diagonal gluon $A_\mu^{\pm}(x)$  
behaves as a massive field with a large mass about 1 GeV 
for $r \ge 0.2 {\rm fm}$ in the MA gauge. 
\item
From the fitting analysis of the lattice QCD data with $r \ge 0.2 {\rm fm}$,  
the off-diagonal gluon mass is evaluated as 
$M_{\rm off} \simeq 1.2~{\rm GeV}$ in the MA gauge.
\end{enumerate}
We perform also the mass measurement of 
off-diagonal gluons from the temporal correlation of 
the zero-momentum projected operator $O_\mu^\pm(\tau)$,  
\bea
\Gamma_{\mu\mu}^{+-}(\tau) \equiv 
\langle O_\mu^+(\tau)O_\mu^-(0) \rangle, 
\quad 
O_\mu^\pm(\tau) \equiv \int d{\bf x} \ A_\mu^\pm({\bf x},\tau), 
\eea
in lattice QCD in the MA gauge plus ${\rm U(1)}_3$ Landau gauge.
We find the off-diagonal gluon mass $M_{\rm off} \simeq 1.2 {\rm GeV}$ 
again from the slope of the logarithmic plot of 
$\Gamma_{\mu\mu}^{+-}(\tau)$ in SU(2) lattice QCD with 
$2.3 \le \beta \le 2.35$ with $16^3\times 32$ and $12^3\times 24$.

Thus, {\it off-diagonal gluons $A_\mu^\pm$ acquire a large effective mass  
$M_{\rm off} \simeq 1.2 {\rm GeV}$ in the MA gauge}, which is 
{\it essence of infrared abelian dominance}.$^{4,5,10}$
In the MA gauge, 
due to the large effective mass $M_{\rm off}\simeq 1.2 {\rm GeV}$, 
off-diagonal gluons $A_\mu^\pm$ can propagate only within 
a short range as $r \lsim M_{\rm off}^{-1} \simeq 0.2{\rm fm}$, 
and becomes {\it infrared inactive} like weak bosons in the Standard Model.
Then, in the MA gauge, off-diagonal gluons $A_\mu^\pm$ cannot contribute to 
the infrared NP-QCD, which leads to 
infrared abelian dominance.$^{4,5,7,10,14}$ 

\begin{figure}[hb]
\begin{center}
\includegraphics[height=3.3cm]{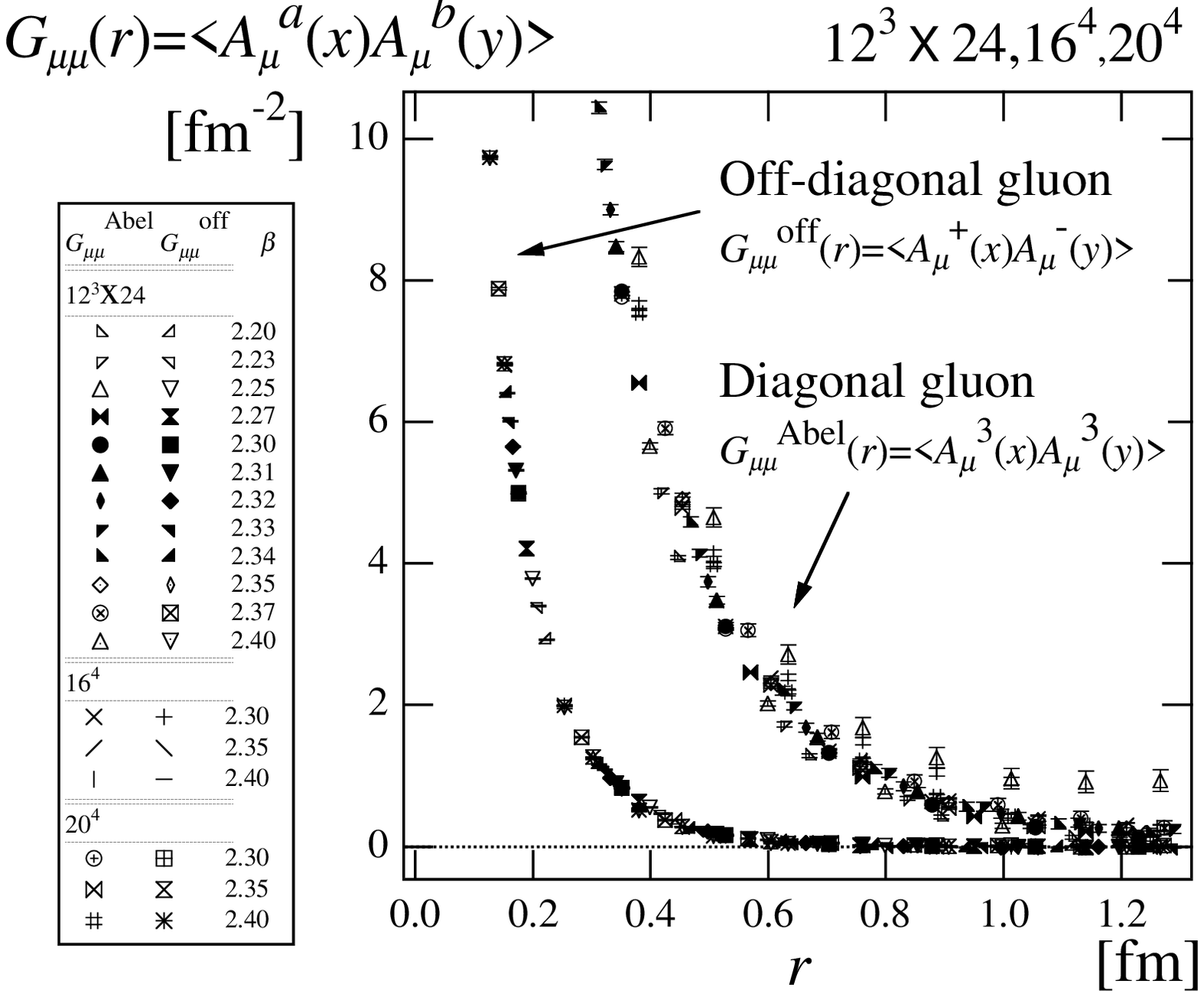}
\includegraphics[height=3.3cm]{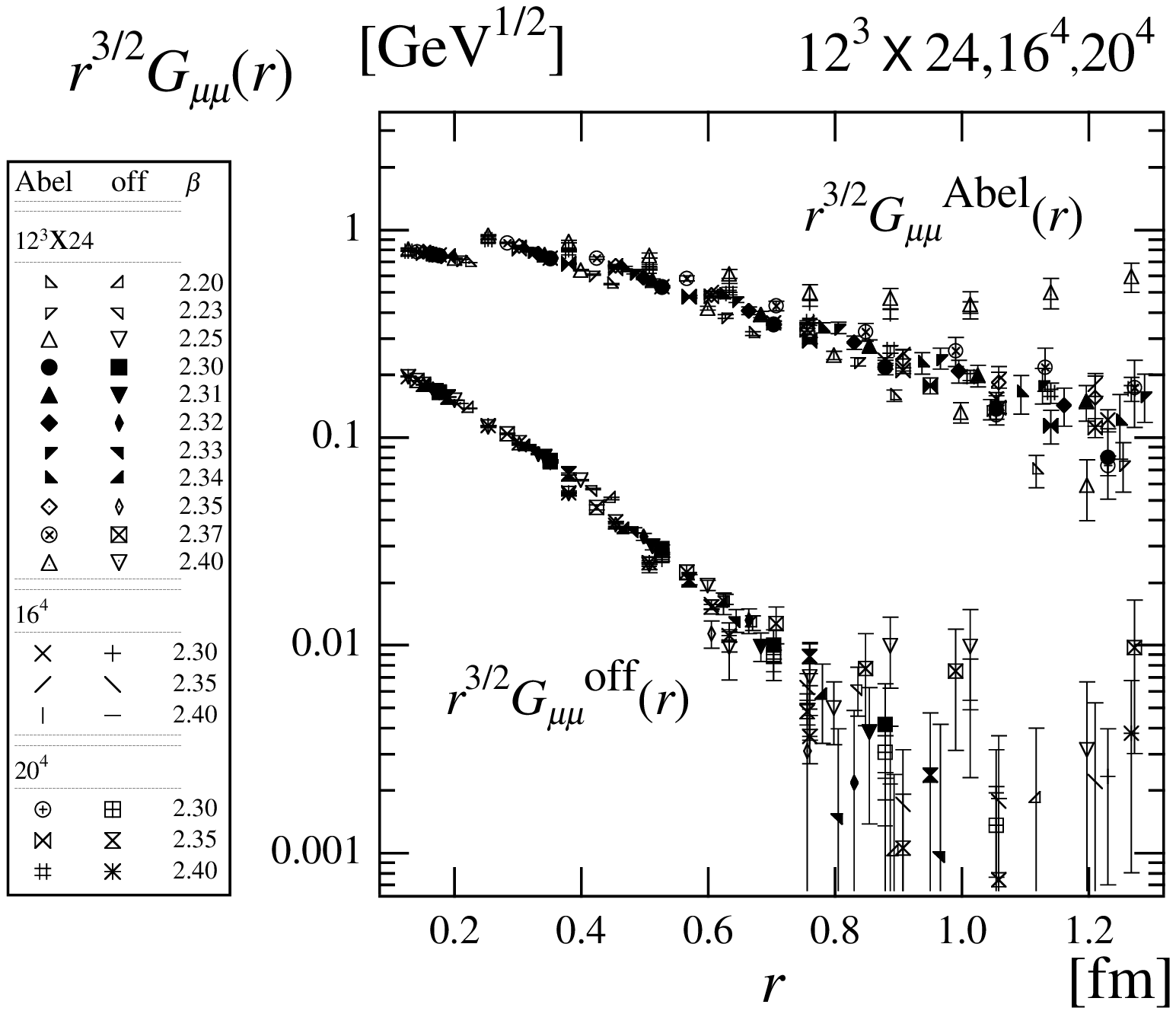}
\includegraphics[height=3.2cm]{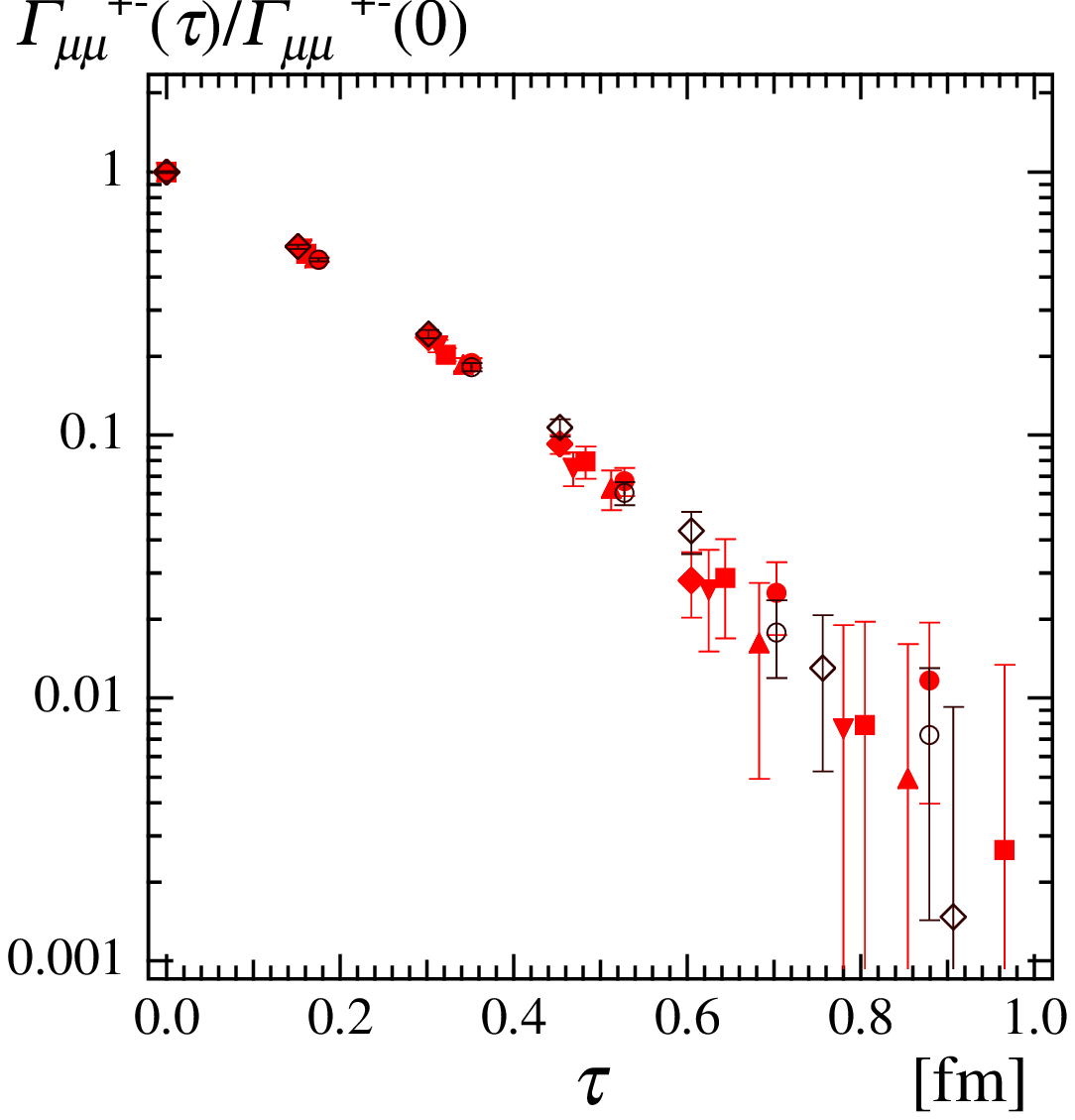}
(a)\hspace{3.6cm}(b)\hspace{3.5cm}(c)
\caption{ (a) The scalar-type gluon propagator 
$G_{\mu \mu }^a(r)$ as the function of the four-dimensional 
distance $r$ in the MA gauge in SU(2) lattice QCD with 
$2.2 \le \beta \le 2.4$ with various sizes 
($12^3 \times 24$, $16^4$, $20^4$). 
(b) The logarithmic plot of $r^{3/2} G_{\mu \mu}^a(r)$ v.s. $r$. 
The off-diagonal gluon propagator behaves as 
the Yukawa-type function, 
$G_{\mu \mu } \sim {\exp(-M_{\rm off}r) \over r^{3/2}}$. 
(c) The logarithmic plot of the temporal correlation 
$
\Gamma_{\mu\mu}^{+-}(\tau) \equiv 
\langle O_\mu^+(\tau)O_\mu^-(0) \rangle 
$
as the function of the temporal distance $\tau$ 
in SU(2) lattice QCD with $2.3 \le \beta \le 2.35$ 
with $16^3 \times 32$ and $12^3 \times 24$. 
From the slope of the dotted lines in (b) and (c), 
the effective mass of the off-diagonal gluon $A_\mu^\pm$ 
is estimated as $M_{\rm off}\simeq 1.2 {\rm GeV}$.} 
\end{center}
\end{figure}

\vspace{-1cm}

\section{Lattice-QCD Evidence of Monopole Condensation}

In the MA gauge, there appears the global network of the 
monopole world-line covering the whole system 
as shown in Fig.4(a), and 
this monopole-current system (the monopole part) 
holds essence of NP-QCD. Using SU(2) lattice QCD, 
we examine the dual Higgs mechanism by 
monopole condensation in the NP-QCD vacuum in the MA gauge.$^{4,5}$ 

Since QCD is described by the ``electric variable'' as 
quarks and gluons, 
the ``electric sector'' of QCD has been well studied with 
the Wilson loop or the inter-quark potential, however, 
the ``magnetic sector'' of QCD is hidden and still unclear. 
To investigate the magnetic sector directly,
it is useful to introduce the ``dual (magnetic) variable'' 
as the {\it dual gluon field} $B_\mu $, which is the dual partner 
of the diagonal gluon 
and directly couples with the magnetic current $k_\mu $.

\begin{figure}[hb]
\begin{center}
\includegraphics[height=3.4cm]{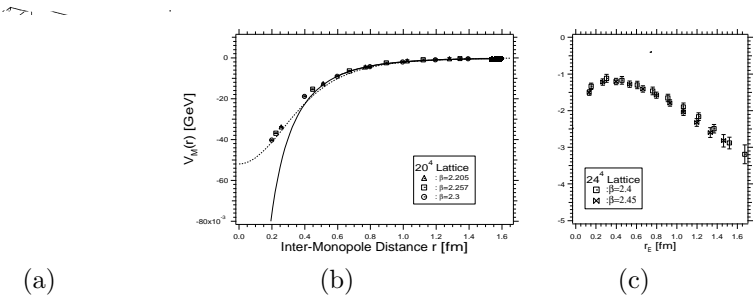}
\includegraphics[height=3.0cm]{Fig4b.EPSF}
\includegraphics[height=3.0cm]{Fig4c.EPSF}
(a)\hspace{3.5cm}(b)\hspace{3.5cm}(c)
\caption{The SU(2) lattice-QCD results in the MA gauge. 
(a) The monopole world-line projected into ${\bf R}^3$ 
on the $16^3 \times 4$ lattice with $\beta =2.2$ 
(the confinement phase). 
There appears a global network of monopole currents covering 
the whole system. 
(b) The inter-monopole potential $V_M(r)$ 
v.s. the 3-dimensional distance $r$ 
in the monopole-current system on the $20^4$ lattice. 
The solid curve denotes the Yukawa potential with $m_B=0.5$GeV. 
The dotted curve denotes the Yukawa-type potential 
including the monopole-size effect. 
(c) The scalar-type dual-gluon correlation 
$\ln (r_E^{3/2} \langle B_\mu(x)B_\mu(y)\rangle_{\rm MA})$ as the 
function of the 4-dimensional Euclidean distance $r_E$ on the 
$24^4$ lattice. The slope corresponds to the dual gluon mass $m_B$.}
\end{center}
\end{figure}

Owing to the absence of the electric current $j_\mu$ 
in the monopole part,
the dual gluon $B_\mu $ can be introduced 
as the regular field satisfying 
$(\partial \land B)_{\mu\nu}={^*\!F}_{\mu\nu}$ 
and the dual Bianchi identity, 
$
{\partial^{\mu}} {^*\!(}\partial \land B)_{\mu\nu}=j_\nu=0.
$
By taking the dual Landau gauge $\partial_\mu B^\mu=0$, 
the field equation is simplified as 
$\partial^2 B_\mu ={\partial_\alpha} {^*\!F}_{\alpha \mu}=k_\mu$, 
and therefore we obtain the dual gluon field $B_\mu$ 
from the monopole current $k_\mu$ as 
\bea
B_\mu (x) = ( \partial^{-2} k_\mu)(x)= -\frac{1}{4\pi^2} 
\int d^4y \frac{k_\mu(y)}{(x-y)^2}. 
\eea
Since the dual gluon $B_\mu $ is massive under monopole condensation, 
we investigate the dual gluon mass $m_B$ 
as the evidence of the dual Higgs mechanism. 

First, we put test magnetic charges in the monopole-current system 
in the MA gauge in SU(2) lattice QCD, and measure the inter-monopole 
potential $V_M(r)$ to get information about monopole condensation. 
Since the dual Higgs mechanism provides the 
{\it screening effect on the magnetic flux}, 
$V_M(r)$ is expected to be short-range Yukawa-type, 
if monopole condensation occurs. 
The potential between the monopole and the anti-monopole 
can be derived as 
\bea
V_{M}(R) = -\lim_{T \rightarrow  \infty} {1 \over T}\ln 
\langle W_D(R,T) \rangle, 
\eea
using {\it the dual Wilson loop} $W_D$ 
as the loop-integral of the dual gluon,$^{4,5}$
\bea
W_D(C) \equiv \exp\{i{e \over 2}\oint_C dx_\mu B^\mu \}=
\exp\{i{e \over 2}\int\!\!\!\int d\sigma_{\mu\nu}{^*\!F}^{\mu\nu}\},
\eea
which is the {\it dual version of the abelian Wilson loop} 
\begin{eqnarray}
W_{\rm Abel}(C) 
\equiv \exp\{i{e \over 2}\oint_C dx_\mu A_3^\mu \} 
=\exp\{i{e \over 2}\int\!\!\!\int d\sigma_{\mu\nu}{F}^{\mu\nu}\}
\end{eqnarray} 
and the test monopole charge is set to be $e/2$. 

In Fig.4(b), we show $V_M(r)$ in the monopole part 
in the MA gauge.$^{4,5}$ 
Except for the short distance, the inter-monopole potential 
is well fitted by the Yukawa potential 
\begin{eqnarray}
V_M(r) = -{{(e/2)}^2 \over 4\pi}{e^{-m_Br} \over r}, 
\end{eqnarray}
and thus the {\it magnetic screening} is observed. 
In the MA gauge, the dual gluon mass is estimated as 
$m_B \simeq {\rm 0.5GeV}$ from the infrared behavior of $V_M(r)$. 

Second, we investigate also the Euclidean scalar-type dual gluon propagator 
$\langle B_\mu(x)B_\mu(y)\rangle_{\rm MA}$ 
as shown in Fig.4(c), 
and estimate the dual gluon mass as $m_B = 0.4 \sim 0.5$ GeV 
from its long-distance behavior.$^{5}$ 

From these two tests, the dual gluon mass is evaluated as 
$m_B=0.4 \sim 0.5$ GeV, 
and this can be regarded as the lattice-QCD evidence 
for the dual Higgs mechanism 
by monopole condensation at the infrared scale.$^{4,5}$ 

To conclude, lattice QCD in the MA gauge exhibits 
{\it infrared abelian dominance} and 
{\it infrared monopole condensation}, 
and therefore the dual Ginzburg-Landau (DGL) theory$^{8,17-23}$ 
can be constructed as the infrared effective theory 
directly based on QCD in the MA gauge. 
(See Fig.5).

\begin{figure}[hb]
\begin{center}
\includegraphics[height=14cm]{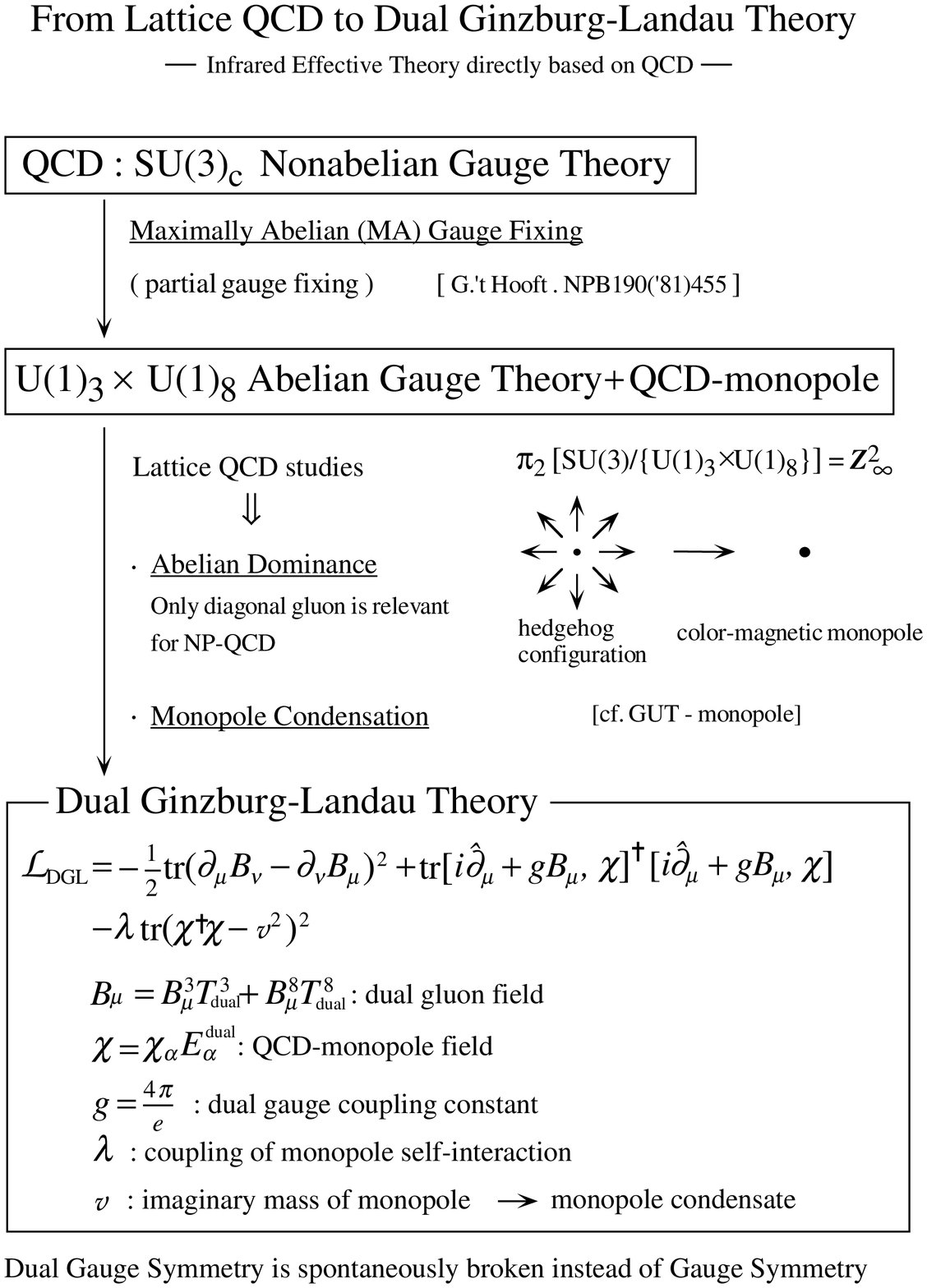}
\caption{
Construction of the dual Ginzburg-Landau (DGL) theory 
from lattice QCD in the maximally abelian gauge. }
\end{center}
\end{figure}


\section{Monopole Structure in terms of the Off-diagonal Gluon}

Let us compare the QCD-monopole with the point-like Dirac monopole. 
There is no point-like monopole in QED, 
because the QED action diverges around the point-like monopole. 
The QCD-monopole also accompanies a large abelian action density, 
however, {\it owing to cancellation with 
the off-diagonal gluon contribution, the total QCD action is 
kept finite even around the QCD-monopole.}$^{3-5}$

To see this, we examine the QCD-monopole structure   
in the MA gauge in terms of the action density 
using SU(2) lattice QCD.$^{3-5}$
From the SU(2) plaquette $P^{\rm SU(2)}_{\mu \nu }(s)$ 
and the abelian plaquette $P^{\rm Abel}_{\mu \nu }(s) 
\in {\rm U(1)}_3  \subset {\rm SU(2)}$, 
we define the ``SU(2) action density'' 
$
S_{\mu \nu }^{\rm SU(2)}(s) 
\equiv 1-{1 \over 2}{\rm tr}P^{\rm SU(2)}_{\mu \nu }(s), 
$
the ``abelian action density'' 
$
S_{\mu \nu }^{\rm Abel}(s) 
\equiv 1-{1 \over 2}{\rm tr}P^{\rm Abel}_{\mu \nu }(s) 
$
and the ``off-diagonal gluon contribution'' 
$
S_{\mu \nu }^{\rm off}(s) 
\equiv S_{\mu \nu }^{\rm SU(2)}(s)-S_{\mu \nu }^{\rm Abel}(s). 
$
In the lattice formalism, 
the monopole current $k_\mu (s)$ is defined on the dual link, 
and there are 6 plaquettes around the monopole. 
Then, we consider the local 
average over the 6 plaquettes around the dual link, 
\begin{eqnarray}
S(s,\mu) \equiv
{1 \over 12} \sum_{\alpha\beta\gamma} \sum_{m=0}^1 
| \varepsilon_{\mu\alpha\beta\gamma} |
S_{\alpha\beta}(s + m \hat \gamma).
\end{eqnarray}
We show in Fig.6(b) the probability distribution of 
the action densities 
$S_{\rm SU(2)}$, $S_{\rm Abel}$ and 
$S_{\rm off}$ around the QCD-monopole in the MA gauge.

\begin{figure}[hb]
\vspace{-0.25cm}
\begin{center}
\includegraphics[height=2.6cm]{Fig6a.EPSF}
\includegraphics[height=2.6cm]{Fig6b.EPSF}
\includegraphics[height=2.9cm]{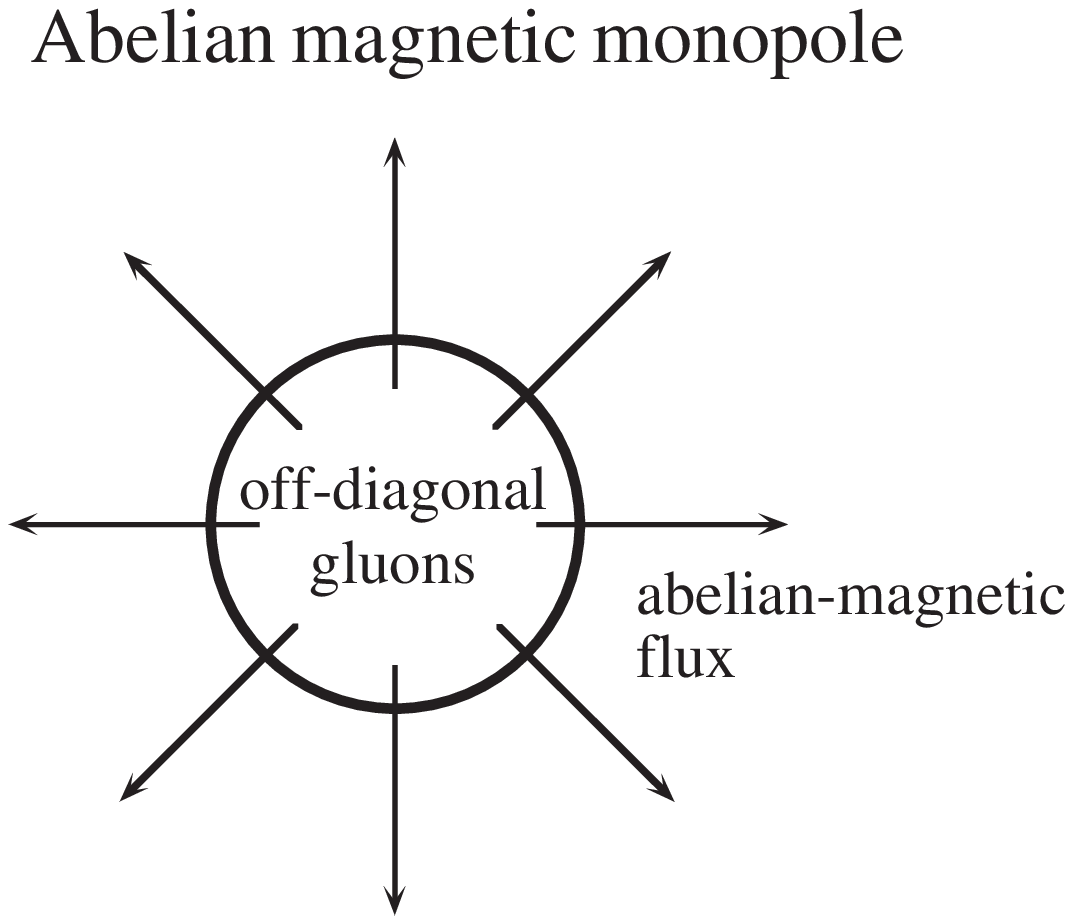}
\end{center}
\begin{center}
(a)\hspace{3cm}(b)\hspace{3cm}(c)
\caption{
(a) The total probability distribution $P(S)$ on the whole lattice 
and (b) the probability distribution $P_k(S)$ around the monopole 
for SU(2) action density $S_{\rm SU(2)}$ (dashed curve), 
abelian action density $S_{\rm Abel}$ (solid curve) 
and off-diagonal gluon contribution $S_{\rm off}$ (dotted curve) 
in the MA gauge at $\beta =2.4$ on $16^4$ lattice. 
Around the QCD-monopole, 
large cancellation between $S_{\rm Abel}$ and  
$S_{\rm off}$ keeps the total QCD-action small. 
(c) The schematic figure for the QCD-monopole structure 
in the MA gauge. 
The QCD-monopole includes a large amount of off-diagonal gluons 
around its center as well as the diagonal gluon.}
\end{center}
\vspace{-0.15cm}
\end{figure}

We summarize the results on the QCD-monopole structure as follows.
\begin{enumerate}
\item
Around the QCD-monopole, 
both the abelian action density $S_{\rm Abel}$ 
and the off-diagonal gluon contribution $S_{\rm off}$ are 
largely fluctuated, and their cancellation keeps 
the total QCD-action density $S_{\rm SU(2)}$ small.
\item
The QCD-monopole has an {\it intrinsic structure 
relating to a large amount of off-diagonal gluons}  
around its center like the 't~Hooft-Polyakov monopole. 
At a large scale, off-diagonal gluons 
inside the QCD-monopole become invisible, 
and QCD-monopoles can be regarded as point-like Dirac monopoles. 
\item
From the concentration of off-diagonal gluons 
around QCD-monopoles in the MA gauge, 
we can naturally understand the 
{\it local correlation between monopoles and instantons}.$^{3-5,11,12,15,16}$
In fact, instantons tend to appear around the monopole world-line 
in the MA gauge, because instantons need 
full SU(2) gluon components for existence.
\end{enumerate}

\section{
Gluonic Higgs and Gauge Invariant Description of MA Projection}

In the MA gauge, the gauge group 
$G \equiv {\rm SU}(N_c)$ is partially fixed into its subgroup 
$H \equiv {\rm U}^{N_c-1}_{\rm local}\times {\rm Weyl}_{N_c}^{\rm global}$, 
and then the gauge invariance becomes unclear.\footnote{ 
In Refs.[3,12], we show a useful {\it gauge-invariance criterion} 
on the operator $O_{\rm MA}$ defined in the MA gauge: 
{\it If $O_{\rm MA}$ defined in the MA gauge is $H$-invariant, 
$O_{\rm MA}$ is also invariant 
under the whole gauge transformation of $G$.} 
}

In this section, we propose a {\it gauge invariant description 
of the MA projection in QCD}. 
Even without explicit use of gauge fixing, we can naturally define the MA 
projection by introducing a ``gluonic Higgs scalar field'' $\phi(x)$.
For a given gluon field configuration $\{ A_\mu(x) \}$, 
we define a gluonic Higgs scalar 
$\vec \phi(x) =\Omega(x) \vec H \Omega^\dagger(x)$ 
with $\Omega(x) \in {\rm SU}(N_c)$ 
so as to minimize 
\bea
R[\vec \phi(\cdot)] \equiv \int d^4x \ {\rm tr} 
\left\{[\hat D_\mu, \vec \phi(x)][\hat D_\mu, \vec \phi(x)]^\dagger \right\}. 
\eea
We summarize the features of this description as follows.$^{3}$

\begin{enumerate}
\item
The gluonic Higgs scalar $\vec \phi(x)$ does not have amplitude degrees 
of freedom but has only color-direction degrees of freedom, and 
$\vec \phi(x)$ corresponds to 
a ``color-direction'' of the nonabelian gauge connection $\hat D_\mu$ 
averaged over $\mu$ at each space-time $x$. 
\item
Through the projection along $\vec \phi(x)$, 
we can extract the abelian U(1)$^{N_c-1}$ sub-gauge-manifold 
which is most close to the original SU($N_c$) gauge manifold. 
This projection is manifestly gauge invariant, and 
is mathematically equivalent to the ordinary MA projection.$^3$ 
\item
Similar to $\hat D_\mu$, the gluonic Higgs scalar $\vec \phi(x)$ 
obeys the adjoint gauge transformation, 
and is diagonalized in the MA gauge. 
Then, monopoles appear at the hedgehog singularities of $\vec \phi(x)$ 
as shown in Fig.7.$^{3,8}$
\item
In this description, 
infrared abelian dominance is interpreted as 
infrared relevance of the gluon mode along 
the color-direction $\vec \phi(x)$, and 
QCD seems similar to a nonabelian Higgs theory. 
\end{enumerate}

\begin{figure}
\begin{center}
\includegraphics[height=3cm]{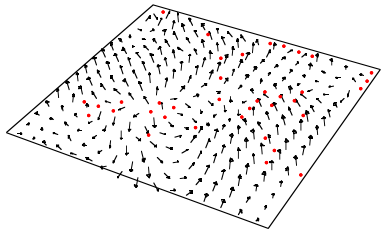}
\caption{
The correlation between the gluonic Higgs scalar field 
$\phi(x)=\phi^a(x)\frac{\tau^a}{2}$ and monopoles denoted by dots 
in SU(2) lattice QCD with $\beta=2.4$ and $16^4$.
The arrow denotes the color direction of 
$(\phi^1(x),\phi^2(x),\phi^3(x))$ in the SU(2) internal space. 
The monopole in the MA gauge appears at the hedgehog singularity of 
the gluonic Higgs scalar $\phi(x)$ in the Landau gauge.}
\end{center}
\end{figure}

\section{The 3Q Ground-State Potential in SU(3) Lattice QCD}

In relation to the color-flux-tube picture for baryons,\cite{CI86,BPV95} 
we study the three-quark (3Q) ground-state potential 
in SU(3)$_c$ lattice QCD at the quenched level.$^{26-28}$
In contrast with a number of studies on the Q-$\bar {\rm Q}$ 
potential using lattice QCD, 
there were only a few preliminary lattice-QCD works for the 
3Q potential.$^{29-31}$ 

To begin with, let us consider the potential form 
in the Q-$\bar{\rm Q}$ and 3Q systems with respect to QCD. 
In the short-distance limit, 
perturbative QCD is applicable and  
the Coulomb-type potential appears 
as the one-gluon-exchange (OGE) result. 
In the long-distance limit at the quenched level,
the flux-tube picture\cite{CI86,BPV95} would be applicable 
from the argument of the strong-coupling limit of QCD\cite{KS75},  
and hence a linear confinement potential 
proportional to the total flux-tube length 
is expected to appear. 
Indeed, lattice QCD results for 
the Q-$\bar {\rm Q}$ ground-state potential are well described by 
\begin{equation}
V_{\rm Q \bar{Q}}(r)=-\frac{A_{\rm Q \bar{Q}}}{r}
+\sigma_{\rm Q \bar{Q}} r+C_{\rm Q \bar{Q}}  
\label{QQpot}
\end{equation}
at the quenched level.  
In fact, $V_{\rm Q \bar{Q}}$ is described by a sum of the short-distance 
OGE result and the long-distance flux-tube result. 

Also for the 3Q ground-state potential $V_{\rm 3Q}$, we try to 
apply this simple picture of the short-distance OGE result 
plus the long-distance flux-tube result. 
Then, the 3Q potential $V_{\rm 3Q}$ is expected to take a form of 
\begin{equation}
V_{\rm 3Q}=-A_{\rm 3Q}\sum_{i<j}\frac1{|{\bf r}_i-{\bf r}_j|}
+\sigma_{\rm 3Q} L_{\rm min}+C_{\rm 3Q}, 
\label{3Qpot}
\end{equation}
with $L_{\rm min}$ the minimal value of 
total length of flux tubes linking the three quarks.
 
Similar to the derivation of the Q-${\bar {\rm Q}}$ potential 
from the Wilson loop, 
the 3Q static potential $V_{\rm 3Q}$ can be derived from the 
3Q Wilson loop $W_{\rm 3Q}$ as~$^{25-31}$  
\bea
V_{\rm 3Q}=-\lim_{T \rightarrow \infty} \frac1T 
\ln \langle W_{\rm 3Q}\rangle, 
\quad 
W_{\rm 3Q} \equiv \frac1{3!}\varepsilon_{abc}\varepsilon_{a'b'c'}
U_1^{aa'} U_2^{bb'} U_3^{cc'} 
\eea
with $U_k \equiv {\rm P}\exp\{ig\int_{\Gamma_k}dx^\mu A_{\mu}(x)\}$ 
($k=1,2,3$) (see Fig.8).  
Here, the 3Q Wilson loop $W_{\rm 3Q}$ is defined in a 
gauge-invariant manner.  
The 3Q gauge-invariant state is generated at $t=0$ 
and is annihilated at $t=T$. For $0 < t < T$, 
the three quarks are spatially fixed in ${\bf R}^3$. 

\begin{figure}
\begin{center}
\includegraphics[height=3.75cm]{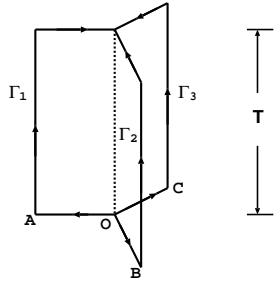}
\caption{The 3Q Wilson loop $W_{\rm 3Q}$.
The 3Q state is generated at $t=0$ and is annihilated at $t=T$. 
The three quarks are spatially fixed in ${\bf R}^3$ for $0 < t < T$.}
\end{center}
\vspace{-0.5cm}
\end{figure}

Physically, the true ground state of the 3Q system 
would be expressed by the flux tubes instead of the strings, 
and the 3Q state which is expressed by the three strings generally includes 
many excited-state components such as flux-tube vibrational modes. 
Since the practical measurement of $\langle W_{\rm 3Q}\rangle$ 
is quite severe for large $T$ in lattice QCD calculations,  
the smearing technique for ground-state enhancement\cite{SMNT00,TMNS00} 
is practically indispensable for the accurate measurement 
of the 3Q ground-state potential $V_{\rm 3Q}$. 

The standard smearing for link-variables\cite{APE87} 
is expressed as the 
iteration of the replacement of the spatial link-variable $U_i (s)$ 
($i=1,2,3$) by the obscured link-variable 
$\bar U_i (s) \in {\rm SU(3)}_c$ which maximizes  
\begin{equation}
{\rm Re} \,\, {\rm tr} \Bigl(
\bar U_i^{\dagger}(s) \Big\{
\alpha U_i(s)+\sum_{\pm, \ j \ne i} 
U_{\pm j}(s)U_i(s \pm \hat j)U_{\pm j}^\dagger (s + \hat i)
\Big\} \Bigr), 
\end{equation}
with a real smearing parameter $\alpha$ and 
$U_{-\mu}(s) \equiv U_{\mu}^{\dagger}(s-\hat \mu)$. 
The $n$-th smeared link-variables $U_\mu^{(n)}(s)$ $(n=1,2,..,N_{\rm smear})$ 
are iteratively defined starting from $U_\mu^{(0)}(s) \equiv U_\mu(s)$ as
\begin{equation}
U_i^{(n)}(s) \equiv \bar U_i^{(n-1)}(s) 
\quad 
(i=1,2,3), 
\qquad 
U_4^{(n)}(s) \equiv U_4(s).
\end{equation}
This smearing procedure keeps 
the gauge covariance of the ``fat'' link-variable $U_\mu^{(n)}(s)$ properly. 
In fact, the gauge invariance of $F(U_\mu^{(n)}(s))$ is ensured 
if $F(U_\mu (s))$ is a gauge-invariant function. 
Since the fat link-variable $U_\mu^{(n)}(s)$ includes a spatial 
extension, the ``spatial line'' expressed with $U_\mu^{(n)}(s)$ physically 
corresponds to a ``flux tube'' with a spatial extension. 
Therefore, if a suitable smearing is done, the ``spatial line'' of the 
fat link-variable is expected to be close to the ground-state flux tube. 
We search reasonable values of the smearing parameter $\alpha$ 
and the iteration number $N_{\rm smear}$ in the lattice QCD calculation. 

\begin{table}[htb]
\tbl{
The best fit coefficients in Eq.(\ref{3Qpot}) for the 3Q potential and 
those in Eq.(\ref{QQpot}) for the Q-$\bar {\rm Q}$ potential in the lattice unit.
\vspace*{1pt}}
{\footnotesize
\newcommand{\m}{\hphantom{$-$}}
\newcommand{\cc}[1]{\multicolumn{1}{c}{#1}}
\renewcommand{\tabcolsep}{2pc} 
\renewcommand{\arraystretch}{1.2} 
\begin{tabular}{@{}llll} \hline \hline
                   & \cc{$\sigma$} & \cc{$A$}      & \cc{$C$}     \\ \hline
${\rm 3Q}$       & $0.1524(28)$  & $0.1331( 66)$  & $0.9182(213)$  \\ 
${\rm Q\bar{Q}}$ & $0.1629(47)$  & $0.2793(116)$ & $0.6203(161)$ \\ 
\hline \hline
\end{tabular} }
\vspace{-0.5cm}
\end{table}
\begin{table}[htb]
\tbl{
Lattice QCD results for the 3Q potential $V_{\rm 3Q}^{\rm lat}$
for 16 patterns of the 3Q system, where the three quarks are put on 
$(i,0,0)$, $(0,j,0)$ and $(0,0,k)$ in ${\bf R}^3$ in the lattice unit.
For each 3Q configuration, $V_{\rm 3Q}^{\rm lat}$ is measured 
from the single-exponential fit 
$\langle W_{\rm 3Q}\rangle=\bar{C}e^{-V_{\rm 3Q}T}$ 
in the range of $T$ listed at the fourth column. 
The statistical errors listed are estimated with the jackknife method,  
and $\chi^2/N_{\rm DF}$ is listed at the fifth column. 
The fitting function $V_{\rm 3Q}^{\rm fit}$ in Eq.(\ref{3Qpot}) with 
the best fitting parameters in Table~1 is also added. 
\vspace*{1pt}}
{\footnotesize
\newcommand{\m}{\hphantom{$-$}}
\newcommand{\cc}[1]{\multicolumn{1}{c}{#1}}
\begin{tabular}{ c c c c c c c } \hline\hline
$(i, j, k)$ & $V_{\rm 3Q}^{\rm lat}$ & \lower.4ex\hbox{$\bar{C}$} &  
fit range & $\chi^2/N_{\rm DF}$
& $V_{\rm 3Q}^{\rm fit}$ & $V^{\rm lat}_{\rm 3Q}-V^{\rm fit}_{\rm 3Q}$ \\
\hline
$(0, 1, 1)$ &0.8457( 38)& 0.9338(173)& 5 --10 & $0.062$ & 0.8524& $-$0.0067 \\ 
$(0, 1, 2)$ &1.0973( 43)& 0.9295(161)& 4 -- 8 & $0.163$ & 1.1025& $-$0.0052 \\ 
$(0, 1, 3)$ &1.2929( 41)& 0.8987(110)& 3 -- 7 & $0.255$ & 1.2929&  \m0.0000 \\
$(0, 2, 2)$ &1.3158( 44)& 0.9151(120)& 3 -- 6 & $0.053$ & 1.3270& $-$0.0112 \\ 
$(0, 2, 3)$ &1.5040( 63)& 0.9041(170)& 3 -- 6 & $0.123$ & 1.5076& $-$0.0036 \\ 
$(0, 3, 3)$ &1.6756( 43)& 0.8718( 73)& 2 -- 5 & $0.572$ & 1.6815& $-$0.0059 \\ 
$(1, 1, 1)$ &1.0238( 40)& 0.9345(149)& 4 -- 8 & $0.369$ & 1.0092&  \m0.0146 \\ 
$(1, 1, 2)$ &1.2185( 62)& 0.9067(228)& 4 -- 8 & $0.352$ & 1.2151&  \m0.0034 \\ 
$(1, 1, 3)$ &1.4161( 49)& 0.9297(135)& 3 -- 7 & $0.842$ & 1.3964&  \m0.0197 \\ 
$(1, 2, 2)$ &1.3866( 48)& 0.9012(127)& 3 -- 7 & $0.215$ & 1.3895& $-$0.0029 \\ 
$(1, 2, 3)$ &1.5594( 63)& 0.8880(165)& 3 -- 6 & $0.068$ & 1.5588&  \m0.0006 \\ 
$(1, 3, 3)$ &1.7145( 43)& 0.8553( 76)& 2 -- 6 & $0.412$ & 1.7202& $-$0.0057 \\ 
$(2, 2, 2)$ &1.5234( 37)& 0.8925( 65)& 2 -- 5 & $0.689$ & 1.5238& $-$0.0004 \\ 
$(2, 2, 3)$ &1.6750(118)& 0.8627(298)& 3 -- 6 & $0.115$ & 1.6763& $-$0.0013 \\ 
$(2, 3, 3)$ &1.8239( 56)& 0.8443( 90)& 2 -- 5 & $0.132$ & 1.8175&  \m0.0064 \\
$(3, 3, 3)$ &1.9607( 93)& 0.8197(154)& 2 -- 5 & $0.000$ & 1.9442&  \m0.0165 \\ 
\hline\hline
\end{tabular} }
\end{table}

We show lattice QCD results for the 3Q ground-state potential. 
We generate 210 gauge configurations using SU(3)$_c$ lattice QCD 
Monte-Carlo simulation with the standard action with 
$\beta=5.7$ and $12^{3} \times 24$ at the quenched level.
The lattice spacing $a \simeq 0.19 \,{\rm fm}$ is determined 
so as to reproduce the string tension as $\sigma$=0.89 GeV/fm
in the Q-$\bar{\rm Q}$ potential $V_{\rm Q \bar{Q}}$.
Here, the pseudo-heat-bath algorithm is used for updating, 
and the gauge configurations are taken every 500 sweeps 
after a thermalization of 5000 sweeps. 

We measure the 3Q ground-state potential $V_{\rm 3Q}$ 
using the smearing technique, and compare the lattice data with 
the theoretical form of Eq.(\ref{3Qpot}).
Owing to the smearing with $\alpha$=2.3 and $N_{\rm smear}$=12, 
the ground-state component is largely enhanced,  
and therefore the 3Q Wilson loop $\langle W_{\rm 3Q} \rangle$ 
composed with the smeared link-variable exhibits 
a single-exponential behavior as 
$\langle W_{\rm 3Q} \rangle \simeq e^{-V_{\rm 3Q}T}$ 
even for a small value of $T$. 
For each 3Q configuration, we extract $V_{\rm 3Q}^{\rm lat}$
from the least squares fit with the single-exponential form
\begin{equation}
\langle W_{\rm 3Q}\rangle =\bar{C}e^{-V_{\rm 3Q}T}
\label{EXfit}
\end{equation}
in the range of $T_{\rm min}\leq T\leq T_{\rm max}$ listed in Table~2.
Here, the fit range of $T$ is chosen such that the stability of the
``effective mass'' 
\bea
V(T)\equiv \ln\{\langle W_{\rm 3Q}(T) \rangle /
\langle W_{\rm 3Q}(T+1)\rangle\}
\eea
is observed in the range of $T_{\rm min}\leq T\leq T_{\rm max}-1$.

For each 3Q configuration, 
we summarize the lattice QCD data $V_{\rm 3Q}^{\rm lat}$ 
as well as the prefactor $\bar{C}$ in Eq.(\ref{EXfit}), 
fit range of $T$ and $\chi^2/N_{\rm DF}$ in Table~2. 
The statistical error of $V_{\rm 3Q}^{\rm lat}$ is estimated 
with the jackknife method. 
We find a large ground-state overlap as $\bar{C} > 0.8$ 
for all 3Q configurations.

Now, we consider the potential form of $V_{\rm 3Q}$. 
We show in Table~1 the best fit parameters in Eq.(\ref{3Qpot}) 
for $V_{\rm 3Q}$.
We compare in Table~2 the lattice QCD data $V_{\rm 3Q}^{\rm lat}$ 
with the fitting function $V_{\rm 3Q}^{\rm fit}$ 
in Eq.(\ref{3Qpot}) with the best fit parameters. 
{\it The three-quark ground-state potential $V_{\rm 3Q}$ is 
well described by Eq.(\ref{3Qpot}) with accuracy better than a few \%,}
although $\chi^2/N_{\rm DF}=3.76$ seems relatively large, 
which may reflect a systematic error on the finite lattice spacing.   
(The fitting with $\Delta$-type flux-tube ansatz 
suggested in Refs.[29,31,34] is rather worse and shows 
unacceptably large $\chi^2/N_{\rm DF}=10.1$ even for the best fit.) 

By comparing the coefficients 
$(\sigma_{\rm 3Q}, A_{\rm 3Q})$ with 
$(\sigma_{\rm Q\bar{Q}}, A_{\rm Q\bar{Q}})$ in Table~1,  
we find a {\it universal feature of the string tension,} 
$\sigma_{\rm 3Q} \simeq \sigma_{\rm Q\bar{Q}}$,  
and the {\it OGE result for the Coulomb coefficient,} 
$A_{\rm 3Q} \simeq \frac12 A_{\rm Q\bar{Q}}$. 

\section*{Acknowledgments}
The authors would like to thank Professor 
Yoichiro~Nambu for his useful suggestions.
The lattice calculations were performed on NEC-SX4
at Osaka University.

\end{document}